\documentclass[aps,prl,reprint,superscriptaddress, showpacs]{revtex4-1}
\usepackage{graphicx}
\usepackage{amsmath}
\usepackage{amssymb}
\usepackage{color}

\begin{document}


\title{Fine structure of the lowest Landau level in suspended trilayer graphene}



\author{H.~J.~van Elferen}
\affiliation{High Field Magnet Laboratory and Institute for Molecules and Materials,
Radboud University Nijmegen, Toernooiveld 7, 6525 ED Nijmegen, The Netherlands}

\author{A.~Veligura}
\affiliation{Physics of Nanodevices, Zernike Institute for Advanced Materials,
University of Groningen, Nijenborgh 4, 9747 AG Groningen, The Netherlands}

\author{N.~Tombros}
\affiliation{Physics of Nanodevices, Zernike Institute for Advanced Materials,
University of Groningen, Nijenborgh 4, 9747 AG Groningen, The Netherlands}

\author{E.~V.~Kurganova}
\affiliation{High Field Magnet Laboratory and Institute for Molecules and Materials,
Radboud University Nijmegen, Toernooiveld 7, 6525 ED Nijmegen, The Netherlands}

\author{B.~J.~van Wees }
\affiliation{Physics of Nanodevices, Zernike Institute for Advanced Materials,
University of Groningen, Nijenborgh 4, 9747 AG Groningen, The Netherlands}

\author{J.~C.~Maan}
\affiliation{High Field Magnet Laboratory and Institute for Molecules and Materials,
Radboud University Nijmegen, Toernooiveld 7, 6525 ED Nijmegen, The Netherlands}
\author{ U.~Zeitler}
\email[]{U.Zeitler@science.ru.nl}
\affiliation{High Field Magnet Laboratory and Institute for Molecules and Materials,
Radboud University Nijmegen, Toernooiveld 7, 6525 ED Nijmegen, The Netherlands}
\email{u.zeitler@science.ru.nl}

\begin{abstract}
Magneto-transport experiments on ABC-stacked suspended trilayer graphene reveal a complete splitting of the twelve-fold degenerated lowest Landau level, and, in particular, the opening of an exchange-driven  gap at the charge
neutrality point. A quantitative analysis of distinctness of the quantum Hall plateaus as a function of field yields a hierarchy of the filling factors:  $\nu=6, 4$, and 0 are the most pronounced, followed by $\nu=3$, and finally
$\nu=1,$ 2 and 5. Apart from the appearance of a  $\nu=4$ state, which is probably caused by a layer asymmetry, this sequence is in agreement with Hund's rules for ABC-stacked trilayer graphene.
\end{abstract}

\pacs{73.22.Pr,73.43.-f,71.70.Di,73.43.Qt}


\maketitle



The unconventional quantum Hall effects observed in single-layer graphene (SLG) \cite{Novoselov_Diracfermions, Kim-QHE},
bilayer graphene (BLG)~\cite{Novoselov_UQHE}, and trilayer graphene (TLG)~\cite{Taychatanapat,Zhang1,Bao4,Kumar1,Jhang1}
are a hallmark for the relativistic bandstructure in this intriguing material.
Considering only nearest-neighbor interactions, the Landau-level spectrum
of all these forms of $N$-layer graphene  can be described by a $4N$-fold degenerate zero-energy level,
shared equally between electrons and holes, and 4-fold degenerate higher Landau levels for electrons and holes separately~\cite{Mccannmultilayer,Guineamultilayer,Koshino2}.

In a magnetic field, exchange effects and the Zeeman splitting can lift this degeneracy~\cite{Ezawa07}.\
However, the mobility in standard samples deposited on a SiO$_2$-substrate is
in general too low in order to resolve such effects. Only when replacing the SiO$_2$-substrate
by e.g.~hexagonal boron nitride (hBN)\cite{Dean1,Dean2} or
by fully suspending the device from the substrate\cite{Bolotin1,Tombros1} the mobility becomes high enough to
completely resolve the fine structure of the lowest Landau level.

In this paper we present magneto-transport experiments
on a suspended ABC-stacked TLG sample. This system is known to display an unconventional QHE
with a twelve-fold degenerate lowest Landau level and Berry phase of 3$\pi$.~\cite{Zhang1,Bao4,Kumar1,Jhang1}
We show that a quantizing magnetic field fully lifts this twelve-fold degeneracy.
Furthermore, by measuring the strengths of quantum Hall plateaus quantitatively,
we established a hierarchic order of the related filling factors: $\nu=6, 4$, and 0 are the most pronounced,
followed by $\nu=3$, and finally $\nu=1, 2$, and 5.

\begin{figure}[t]
\centering
\includegraphics[width=0.95\linewidth,angle=0]{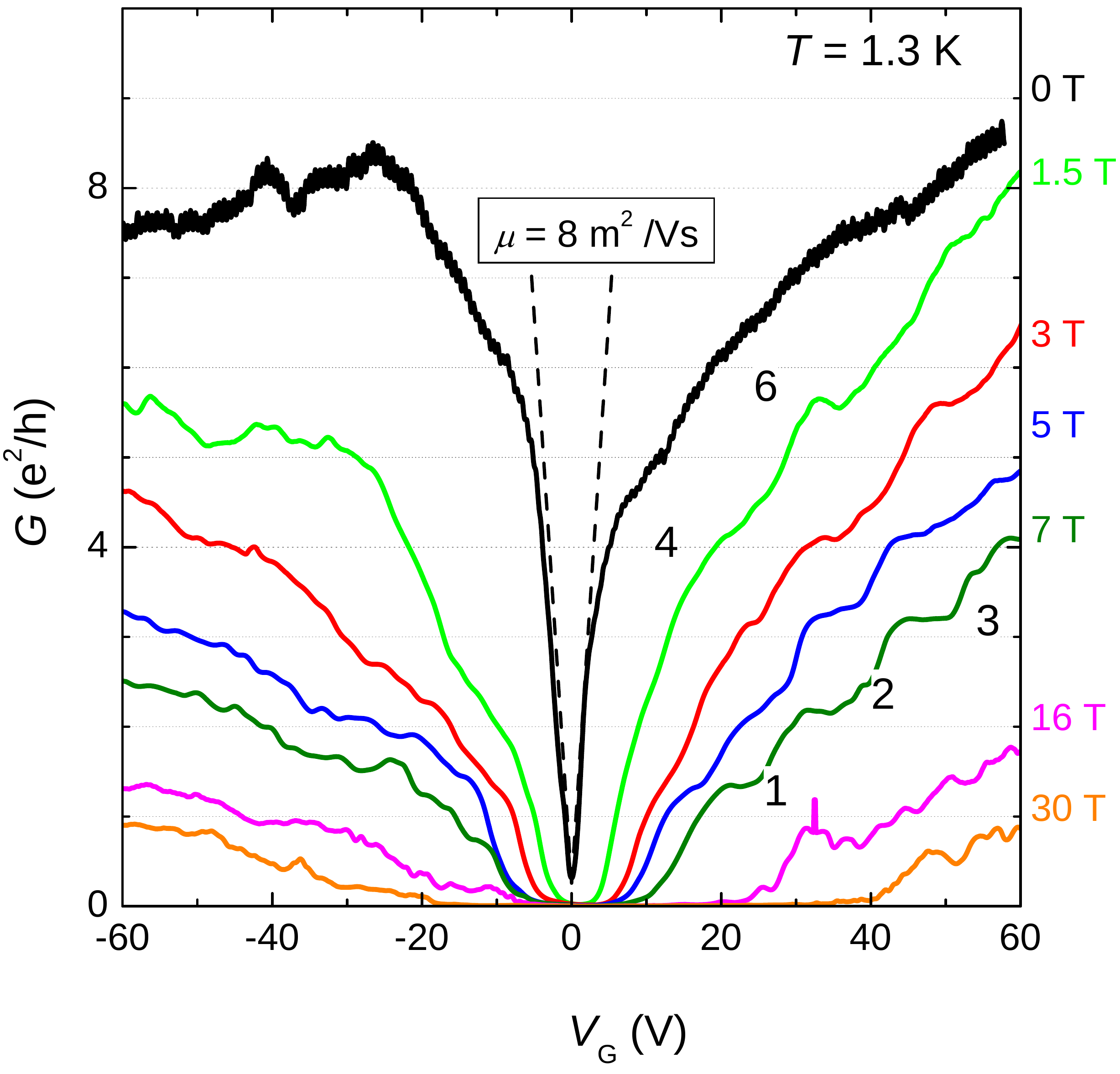}
\caption{
Conductance traces at $T=$~1.3~K for magnetic fields between 0 and 30~T. The dashed line through the 0~T data is used to estimate the device mobility. The numbers indicate quantization of $G$ in integer units of $e^2/h$.
}
\label{Fig1_QHEtrilayer}
\end{figure}

We have prepared a suspended TLG sample using an acid free method.\cite{Tombros1}
Following standard techniques,~\cite{Novoselov_firstgraphene} we first exfoliated flakes from highly oriented pyrolytic graphite and deposited them on a Si/SiO$_2$ substrate covered with a 1.15~$\mu$m thick LOR-A resist layer. The
TLG-flake was then identified by its thickness measured through its optical contrast.~\cite{Blake1} Subsequently, two electron beam lithography steps were performed in order to contact the flake with Ti-Au contacts and to remove
part of the LOR-A below the graphene flake. The resulting device is freely suspended across a trench formed in the LOR-A with two metallic contacts on each side. Carriers in the sample are induced by applying a back-gate voltage
$V_G$ on the highly $n$-doped Si-wafer yielding a carrier concentration $n = \alpha V_G$. The lever factor $\alpha  \approx 1 \times 10^{14}$~m$^{-2}$V$^{-1}$ is determined experimentally from the positions of the filling factors
in Fig.~\ref{Fig1_QHEtrilayer} and agrees within a factor of two with that deduced from the geometric gate capacitance of the device before annealing.

Measurements were performed at low temperatures and high magnetic fields up to 30 T using a low-frequency (1.87~Hz) lock-in technique with an excitation current $I \leq$ 1~nA.
The sample was mounted on an in-situ tilting stage where the angle $\phi$ between the total magnetic field $B_{tot}$ and the perpendicular component $B_\perp=B \cos{(\phi)}$, can be controlled independently. $\phi$  was determined
using the Hall-resistance of a second sample on the same substrate.
The device was slowly cooled down to 4.2~K and current annealed~\cite{Moser1} by applying a dc bias current up to 3~mA. The local annealing resulted into a high quality sample where the charge neutrality point (CNP) is centered
around zero gate voltage.

In Fig.~\ref{Fig1_QHEtrilayer} we show the two-terminal conductance $G$ of our sample as a function of $V_G$ in a perpendicular magnetic field ($\phi=0$). Before calculating the conductance, we have subtracted a constant
background-resistance of 550~$\Omega$ originating from the finite contact- and lead-resistance from the measured two-terminal resistance. Using the slope of the dashed line in the figure, $G = n e \mu w/l$, we estimate a
zero-field mobility $\mu \approx 8$~m$^2$/Vs around the CNP. Here $l$ and $w$ are the length and the width of the sample and we assume that their ratio did not change significantly during annealing. A value of the order of a few
m$^2$/Vs for the mobility is further confirmed by the fact that we start entering the quantum Hall regimes already around 1 T (see below).

At  $B < 3$~T, quantum Hall plateaus at filling factors $\nu=4$ and $\nu=6$ already start to develop.
A further increase of the magnetic field up to 10~T results in the complete lifting of the lowest Landau level
and the formation of quantized Hall plateaus at filling factors $\nu=5$, 3, 2 and 1.

\begin{figure}[t]
\centering
\includegraphics[width=0.95\linewidth,angle=0]{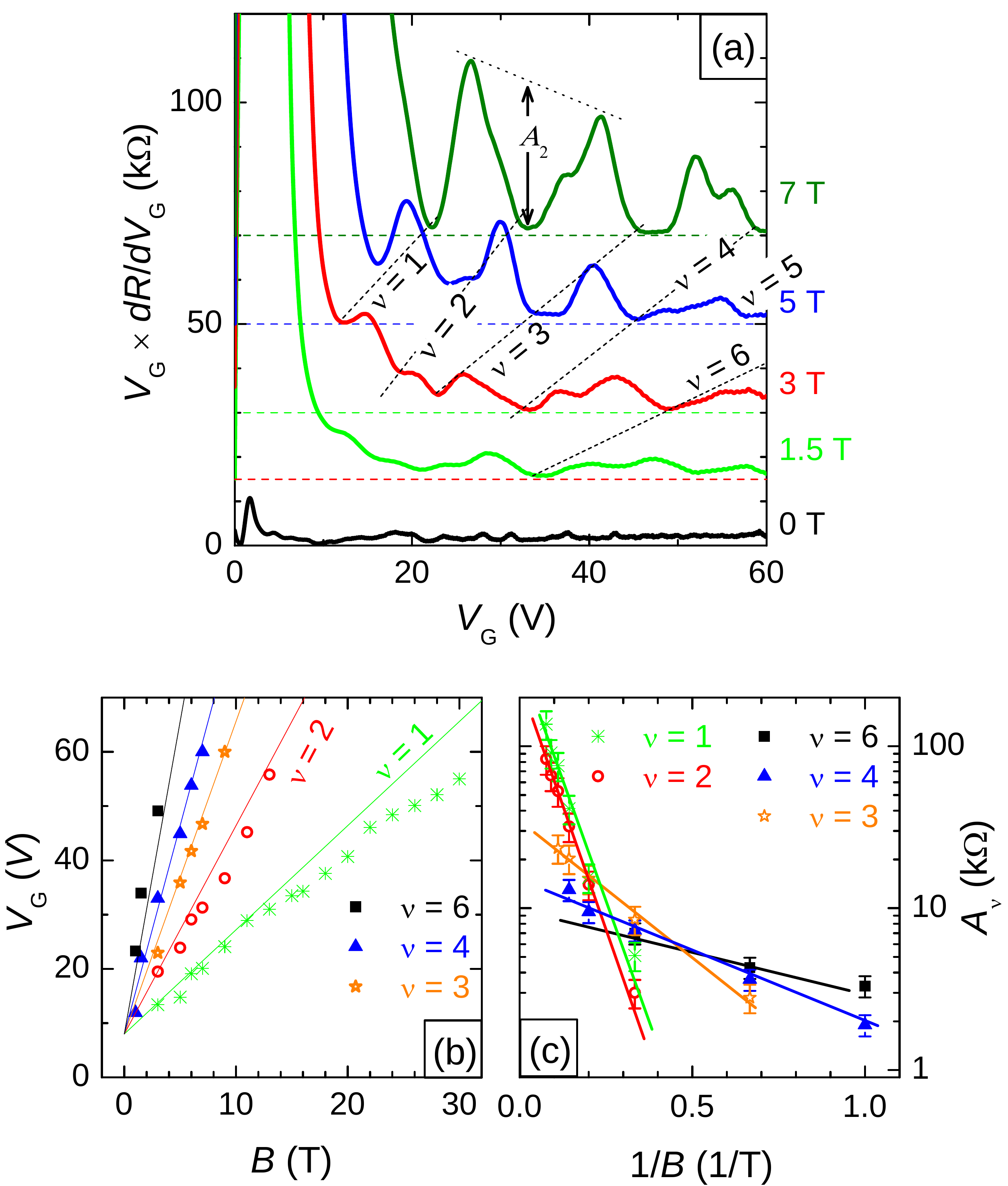}
\caption{
(a) Derivative $D= V_G \times dR/dV_G$ for magnetic fields between 0 and 7 T. Curves are shifted upwards proportional to the magnetic field value. The dashed lines mark the position of the minima in $D$  with the corresponding
filling factor. The arrow illustrates the definition of the oscillation amplitude $A_{\nu}$ at $\nu=2$.\protect{\newline}
(b) Gate-voltage positions of the minima in $D$ as a function of field. The lines indicate the expected linear behavior for the giving filling factors.\protect{\newline}
(c) Oscillation amplitude $A_{\nu}$ as a function of $1/B$ for the different filling factors.}
\label{Fig2_DingleT}
\end{figure}

The conductance $G$ is the inverse of the resistance $R$, which is determined by a combination of the magnetoresistance $R_{xx}$ and the Hall resistance $R_{xy}$ (after subtraction of contact- and leads resistances). Our data for
high concentrations show that $R$  is dominated by $R_{xy}$, indicated by the formation of plateaus in Fig.~\ref{Fig1_QHEtrilayer}. Motivated by the empiric relation $R_{xy} \propto B \times dR_{xx}/dB$,\cite{Chang,Benno} and in
order to accentuate the plateaus more clearly we define therefore a normalized derivative:
\begin{equation}
D = -V_G \frac{dR}{dV_G}
\label{defD}
\end{equation}
which is plotted in Fig.~\ref{Fig2_DingleT}(a) for the measured data in Fig.~\ref{Fig1_QHEtrilayer}.
In this way, plateaus in $R$, originating from $R_{xy}$, result into clear minima at integer filling factors $\nu = ne/hB$ that are related to Shubnikov-de Haas minima in $R_{xx}$.
These minima are well pronounced on the electron side, $V_G  > 0$, dashed lines in Fig.~\ref{Fig2_DingleT}(a).
On the hole side, a changing  background-resistance, possibly originating from less well-annealed parts of the sample,
makes it more hard to distinguish the different plateaus and corresponding minima, though they still remain visible.
Therefore, we will focus our  analysis on the electron side only.

In Fig.~\ref{Fig2_DingleT}(b)
we plot the position of the minima in $D$ as a function of gate voltage (proportional to $n$).
As expected, they show a linear magnetic field dependence. The finite offset at zero field is probably caused by
the persistence of the quantized states down to zero field. \cite{Freitag,Tombros2}

We now focus on the quantitative development of filling factors $\nu =  6, 4, 3, 2,$ and 1 by determining a typical magnetic field $B_0$ at which quantization appears. As can be seen in Fig.~\ref{Fig1_QHEtrilayer}(b), the
oscillation amplitude increases with increasing field until fully developed plateaus appear in $G$. We use a quantitative analysis of the amplitudes $A_{\nu}$  similar to the determination of the Dingle temperature $T_D$ in
Ref.~\onlinecite{Erik1}. This model is based on the Lifshitz-Kosevich formula,\cite{Kosevic} $\Delta R = A_\nu(B,T) \cdot \sin (P/B + \varphi )$, with an oscillation amplitude $A_\nu(B,T)$, a period $P$ and a phase $\varphi$.
Depending on the corresponding gap at a given filling factor $\nu$, the amplitude is different for different $\nu$. $A_\nu(B,T)$ contains a temperature dependent term $R_T$ and a field dependent Dingle term $R_D$. In order to
concentrate on the field dependence alone, we have performed all measurements at a constant temperature $T = 1.3$~K, i.e.~leaving $R_T$ constant for all measurements.

In a regime where two neighboring Landau levels are still overlapping, the Dingle-factor at the oscillation minima scales as $R_B \propto \exp(-B_0/B)$. For higher fields, where the levels are fully separated, $R_D$ saturates and
becomes field independent.
Filling factors with the largest excitation gap appear first at the lowest $B$, while filling factors corresponding to smaller gaps appear at higher $B$. Quantitatively, using the above equation, we define an onset field $B_0$
where 37~\% ($1/e$)  of the maximum oscillation amplitude is reached.

In Fig.~\ref{Fig2_DingleT}(c) we plot the amplitude $A_{\nu}$ (defined as the distance from the oscillation minimum to the average of the two neighboring maxima, see arrow in Fig.~\ref{Fig2_DingleT}(a)), as a function of $1/B$ for
the different filling factors. The solid lines indicate the slope of the data points and determine the values of $B_0$ summarized in Tab.~\ref{table1_suspendedtrilayer}. The $\nu =5$ minimum just starts to appear in the 5~T trace
in Fig.~\ref{Fig2_DingleT}(a); due to the limited gate voltage range we were not able to perform a quantitative analysis of its amplitude $A_5 (B)$.

As stated above, the values of $B_0$ in Tab.~\ref{table1_suspendedtrilayer} describe a hierarchy of the filling factors.  The most pronounced filling factor is found to be $\nu=6$, separating the lowest 12-fold degenerate Landau
level of ABC-stacked TLG from the first Landau level. It confirms that our sample is indeed TLG.\cite{Zhang1,Bao4,Kumar1,Jhang1}
The subsequent filling factors are related to the lifting of the degeneracy of the lowest Landau level: First $\nu=3$ and at higher fields $\nu=$~1, 2 and 5,
a sequence which is in agreement with Hund's rules of ABC-TLG.\cite{Zhangbreakingtrilayer}
However, filling factor $\nu=4$ is also observed experimentally in this sequence whereas it is not predicted by Hund's rules for ABC-TLG.
We attributed its appearance to a layer asymmetry caused by an external electric field of the backgate or local inhomogeneities.\cite{Yuan1}

\begin{figure}[t]
\centering
\includegraphics[width=0.95\linewidth,angle=0]{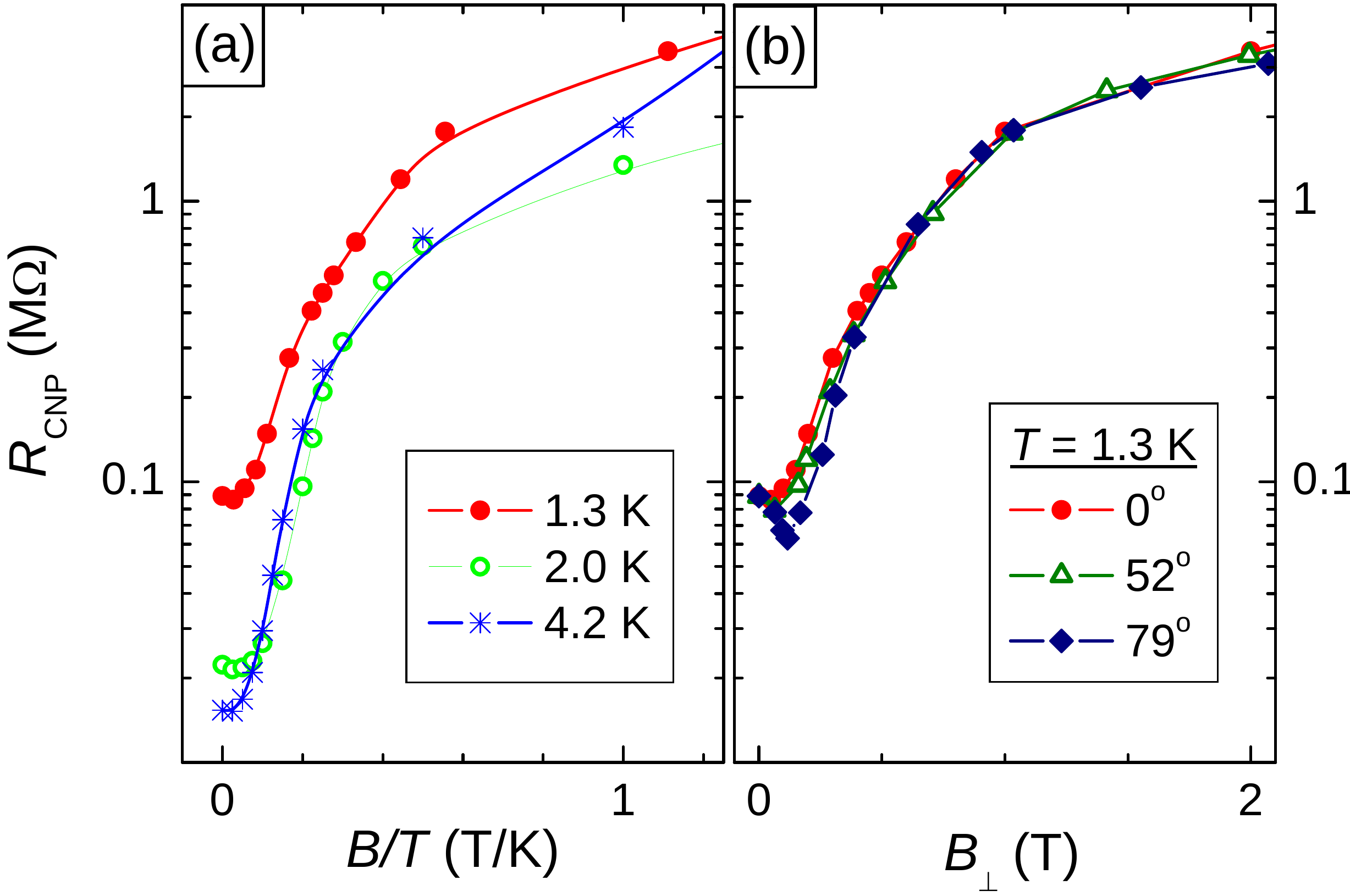}
\caption{
(a) $R_{CNP}$ in a magnetic field perpendicular to the 2DES plotted as a function of $B/T$ for different temperatures.\protect{\newline}
(b) $R_{CNP}$ at 1.3~K in tilted magnetic fields plotted as a function of the perpendicular field component $B_\perp$.
}
\label{Fig3}
\end{figure}

\begin{table}[b]
    \begin{center}
        \begin{tabular}{cccccc}
        \hline
        \hline
        $\nu$ & 6 & 4 & 3 & 2 & 1\\
        \hline
        $B_0$ (T) ~&~ $1.4 \pm 0.5$ ~&~ $2.2 \pm 0.5$ ~&~ $4 \pm 1$ ~&~ $14 \pm 2$ ~&~ $14 \pm 2$\\
        \hline
        \hline
        \end{tabular}
    \end{center}
    \vspace*{-1em}
    \caption{Field values $B_0$ characterizing the strength of the plateaus at $\nu=6$, 4, 3, 2, and 1.}
    \label{table1_suspendedtrilayer}
\end{table}

Additionally, a field-induced gap opens at the CNP, as theoretically expected for ABC-stacked TLG\cite{Zhangbreakingtrilayer} but not for ABA-stacked TLG.\cite{Koshino1} We analyse the nature of this gap in more detail by
focussing  on the diverging resistance at the CNP. Already at zero magnetic field a strong activated temperature dependence is observed (see Fig.~\ref{Fig3}(a)). It can be described by $R_{CNP} \propto \exp{(\Delta_0/k_B T)}$,
with a gap-size $\Delta_0 \approx$~0.38 meV. In a magnetic field, $R_{CNP}$  grows rapidly with increasing $B/T$, suggesting an increase of the relevant gap. For $B/T$ up to 0.5~T/K the roughly exponential behavior of $R_{CNP}$
suggests an Arrhenius-activated transport with a field enhanced gap $\Delta (B) = \Delta_0 + \gamma B$ with $\gamma=1.1$~meV/T.  This value is one order of magnitude larger than the bare Zeeman gap $\Delta=g \mu_B B$ (0.116~meV at
1 T). We therefore suggest that an exchange-enhanced mechanism is responsible for the field-induced gap opening.

The scenario of an exchange-driven gap at the CNP is further supported by experiments in tilted magnetic fields shown in Fig.~\ref{Fig3}(b). The resistance at the CNP is  governed by the perpendicular magnetic field $B_\perp$
alone consistent with a non-spin related mechanism responsible for the gap at the CNP. This leads to an insulating phase of a quantum Hall insulator, as also proposed for single-layer and bilayer graphene.\cite{Jiang1,Ong}

\begin{figure}[t]
\centering
\includegraphics[width=0.95\linewidth,angle=0]{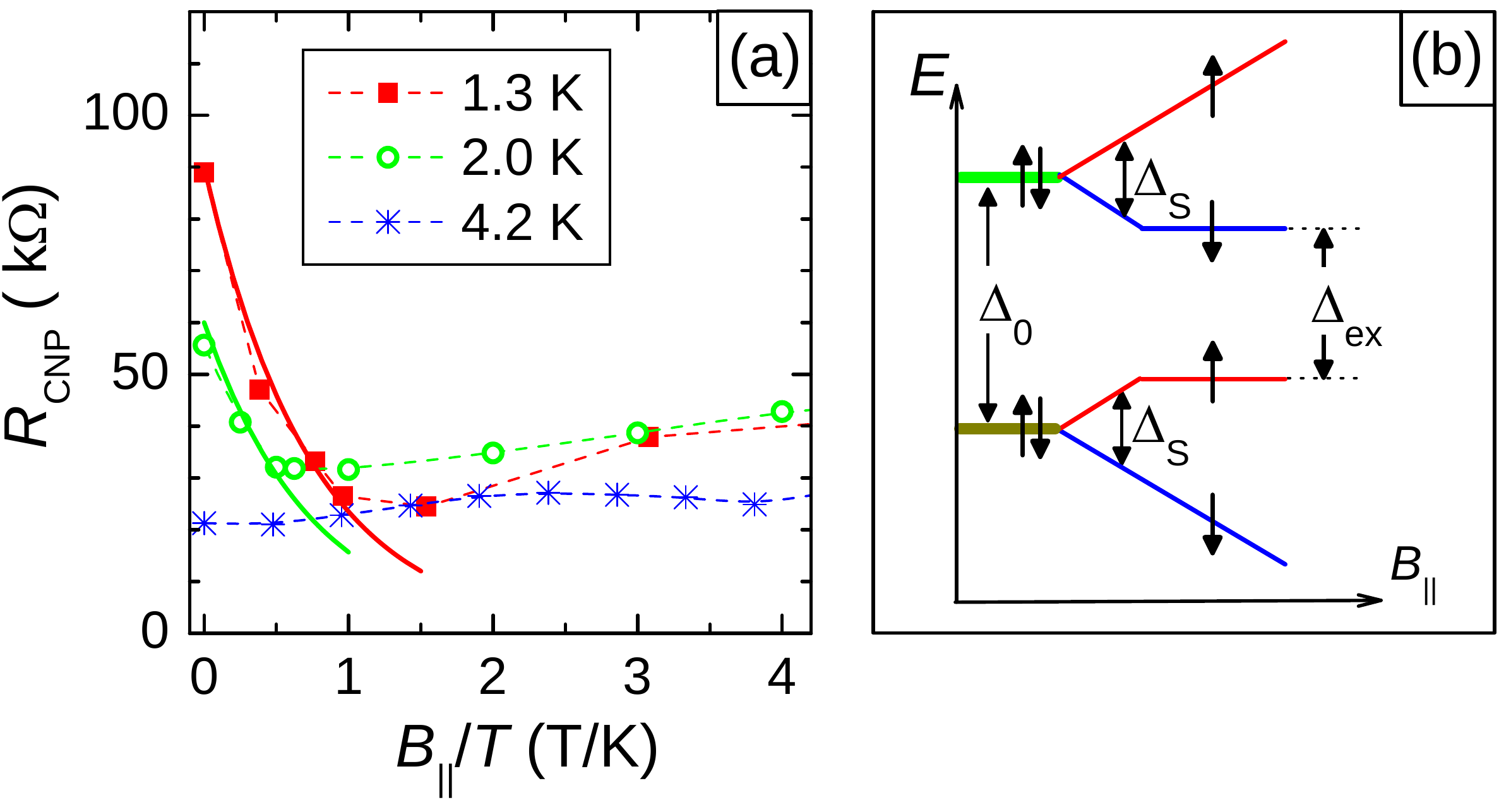}
\caption{(a) Resistance at the CNP in a parallel magnetic field plotted as a function $B/T$. The solid lines show the expected resistance decrease, $R \propto \exp{(-2 \mu_B B /k_B B_\parallel T)}$,  scaling with the bare
Zeeman-energy. (b) Proposed scenario for the behavior in a parallel magnetic field: A gap $\Delta_0$ present at 0~T closes due to spin-splitting in both energy levels. An exchange mechanism prevents further decrease of the gap for
parallel fields above 2~T, a finite gap $\Delta_{ex}$~=~0.25~meV remains.}
\label{Fig4}
\end{figure}

Finally, we demonstrate that the bare Zeeman splitting at the CNP can be observed directly using a parallel magnetic field $B_\parallel$.
As shown in Fig.~\ref{Fig4}(a), the resistance at the CNP decreases as a function of $B_\parallel /k_B T$, similar to a recent observation in a suspended BLG sample.\cite{Veligura1}
This decrease suggests that the gap is suppressed by a magnetic field, $\Delta = \Delta_0 - \Delta_\parallel (B)$.
Assuming a simple Zeeman gap, $\Delta_\parallel (B) =2\mu_B B_\parallel$, leads to a field dependence shown by the solid lines in Fig.~\ref{Fig4}(a). Indeed, the lines follow the experimental data up to  $B_\parallel/T \approx~1$
T/K.
Therefore, we can understand the field induced gap closing at the CNP as a simple Zeeman splitting of the
two spin-degenerate levels above and below the Fermi energy.  For higher fields, the resistance remains constant i.e.~an exchange mechanism prevents the two energy levels to approach and the gap remains open at an approximately
constant value.

In conclusion, magnetotransport experiments on a two-probe suspended ABC-stacked trilayer graphene sample show the full lifting of the twelve-fold degeneracy in the lowest Landau level. Performing a quantitative analysis on the
distinctness of the related quantum Hall plateaus we have determined an order for the appearance of the corresponding filling factors:  $\nu=6, 4, 0$ appear first, followed by $\nu=3$  and finally $\nu=1, 2, 5$. Furthermore, we
have studied the opening of a gap at the CNP. Already at zero magnetic field we observe a gap $\Delta_0=$~0.38~meV, which can be partly closed by the Zeeman effect in a parallel magnetic field. In contrast, in a perpendicular
magnetic field, the gap at $\nu=0$ was shown to increase linearly with field and to be an order of magnitude larger than the bare Zeeman gap. These facts point to a spin unpolarized ground state with an exchange-driven gap.

Part of this work has been supported by the Stichting Fundamenteel Onderzoek der Materie (FOM), with financial support from the Nederlandse Organisatie voor Wetenschappelijk Onderzoek (NWO). We also thank Zernike Institute for
Advanced Materials  and Nanoned for financial support.


%

\end{document}